\documentclass[prb, twocolumn, superscriptaddress]{revtex4-2}
\usepackage{bm, amsmath, amsfonts, amssymb, mathtools, ascmac, braket}
\usepackage{multirow}
\usepackage{graphicx}
\usepackage{float, xcolor}
\usepackage{physics}
\usepackage{comment}

\usepackage[
pagebackref=false,
colorlinks=true,
linkcolor=black,
urlcolor=black,
filecolor=black,
citecolor=blue,
pdfstartview=FitV,
pdftitle={},
pdfauthor={},
pdfsubject={},
pdfkeywords={},
pdfpagemode=None,
bookmarksopen=true
]{hyperref}

\newcommand{\ii}{\text{i}}

\newcommand{\revise}[1]{{\textcolor{black}{#1}}}

\newcommand{\imp}[1]{{\textcolor{black}{#1}}}

\begin{document}

\title{Interacting Electronic Topology of Nonlocal Crystals}

\author{Shu Hamanaka}
\email{hamanaka.shu.45p@st.kyoto-u.ac.jp}
\affiliation{Department of Physics, Kyoto University, Kyoto 606-8502, Japan}

\author{Martina O. Soldini}
\email{martina.soldini@physik.uzh.ch}
\affiliation{University of Zurich, Winterthurerstrasse 190, 8057 Zurich, Switzerland}

\author{Tsuneya Yoshida}
\email{yoshida.tsuneya.2z@kyoto-u.ac.jp}
\affiliation{Department of Physics, Kyoto University, Kyoto 606-8502, Japan}

\author{Titus Neupert}
\email{titus.neupert@physik.uzh.ch}
\affiliation{University of Zurich, Winterthurerstrasse 190, 8057 Zurich, Switzerland}

\date{\today}

\begin{abstract}
Nonlocal crystals are systems with translational symmetry but arbitrary-range couplings or interactions between degrees of freedom.
\revise{We argue that such systems can realize symmetry-allowed but locality-forbidden ground-state symmetry eigenvalues.
This is demonstrated by constructing an exactly solvable fermionic model in a one-dimensional translationally invariant setting in symmetry class AII with inversion symmetry, using a Hatsugai-Kohmoto-type model.
The resulting unique gapped ground state exhibits the size-independent inversion eigenvalue $-1$, which is allowed from the fixed-size zero-dimensional viewpoint but forbidden in both noninteracting band insulators and local crystalline fermionic symmetry-protected topological phases.
Our work establishes Hatsugai-Kohmoto-type nonlocality as a concrete route to realizing symmetry sectors that are inaccessible in such local phases.
}
\end{abstract}

\maketitle


\section{Introduction}
The conventional definition of an insulator is a gapped state of matter which can be described in terms of localized degrees of freedom~\cite{Kohn-PR-64}. This notion was supplanted by the concept of topological insulators~\cite{Hasan-Kane,Qi-Zhang}, where nonlocal features arise despite the insulating bulk. Key examples are Chern insulators~\cite{Haldane-PRL-88,QWZ-PRB-06,*Qi-PRB-08}, where topology presents an obstruction to exponential localization of symmetric Wannier functions~\cite{Thouless-84,Brouder-PRL-07,Soluyanov-PRB-11}. 
Such topological phases are encompassed by the classification of fermionic symmetry-protected topological states (FSPTs)~\cite{Schnyder-PRB-06,Kitaev-09,Ryu-NJP-10,Fidkowski-1,*Fidkowski-2,Wang-PRB-14, Song-PRB-17, Chen-arxiv-2024} which have gapped, short-range entangled, and nondegenerate ground states, as well as gapless edge modes that are protected by symmetry. If the protecting symmetry is a symmetry of the lattice, these are crystalline FSPTs~\cite{Fu-PRB-07,Fu-PRL-11,Morimoto-Furusaki,Shiozaki-Sato, Song-PRX-17,Huang-PRB-17,Thorngren-PRX-18,Else-PRB-19}.

Despite their nonlocal properties, FSPTs -- and topological phases in general -- are born of local Hamiltonians. Locality, along with the adiabatic theorem, is the key pillar for classifying topological phases in different dimensions. There are two reasons for this: (1) their key characteristics, edge states and quantized topological response functions, are not protected with infinite-range interactions, (2) the very notion of dimensionality seems lost as long-range couplings render any system effectively zero-dimensional.
Previous works explored the resilience of FSPTs with local Hamiltonians under the addition of long-range interactions~\cite{Viyuela-PRB-16,Dutta-PRB-17,Solfanelli-JHP-23}, and asked whether adding long-range terms enables novel physics~\cite{Defenu-RMP-23}.

\revise{
In this paper, we study how relaxing locality changes the set of symmetry
quantum numbers that can be realized by gapped translationally invariant ground states. 
We refer to translationally invariant systems with arbitrary-range couplings or interactions as \emph{nonlocal crystals}.
At fixed system size, such systems can be viewed as zero-dimensional
quantum systems in which translation acts as an internal symmetry.
Thus, the relevant question is not whether nonlocality evades the
zero-dimensional viewpoint, but which symmetry-allowed quantum numbers are forbidden by locality in one-dimensional crystalline phases and can nevertheless be realized by nonlocal crystalline Hamiltonians.
}

\revise{
Specifically, we consider electrons arranged on a one-dimensional lattice with spinful time-reversal symmetry (TRS), charge conservation, lattice translations, and inversion symmetry. 
In this setting, we show that an infinite-range interaction realizes a unique gapped ground state with the size-independent inversion eigenvalue \(I(L)=-1\), where $L$ is the linear system size.
This inversion eigenvalue is allowed by the fixed-size zero-dimensional symmetry representation theory, but is forbidden in
both noninteracting band insulators and local interacting
one-dimensional crystalline FSPTs. We construct an exactly solvable representative of this
symmetry-allowed but locality-forbidden sector, inspired by the model originally proposed by Hatsugai and Kohmoto (HK)~\cite{HK-92-JPSJ}, which has been recently extended to superconductors~\cite{Phillips-NPhys-20,Zhao-PRB-22}.
}

\revise{
We also discuss how the comparison between fixed-size
zero-dimensional symmetry representation theory, noninteracting band
theory, and local interacting crystalline phases extends to higher
dimensions, and comment on possible platforms for realizing
Hatsugai-Kohmoto-type nonlocal interactions.
}

\revise{
The paper is organized as follows.
In Sec.~\ref{sec:gs inv ev}, we introduce the setup and the
ground-state inversion eigenvalue \(I(L)\) used throughout this work.
We then enumerate the possible values of \(I(L)\) in three settings:
fixed-size zero-dimensional symmetry representation theory, noninteracting band insulators, and local interacting one-dimensional crystalline FSPTs.
This identifies the size-independent value \(I(L)=-1\) as
symmetry-allowed at fixed size but forbidden by locality in
one-dimensional crystalline FSPTs.
In Sec.~\ref{sec:HK}, we construct an exactly solvable
Hatsugai-Kohmoto-type nonlocal model and show that it realizes this
value as a unique gapped ground state.
In Sec.~\ref{sec:berry}, we use Resta's formula to evaluate the
many-body center of mass of the constructed ground state under periodic boundary conditions.
In Sec.~\ref{sec:generalization}, we discuss extensions to higher
dimensions.
In Sec.~\ref{sec:discussion}, we summarize the results and provide some outlook.
In Appendix~\ref{asec:hkreal}, we describe in detail how locality is
imposed on the HK-type interaction.
In Appendix~\ref{asec:classification}, we summarize the real-space construction classification of one-dimensional class-AII crystalline FSPTs with inversion and translation symmetries. 
In Appendix~\ref{asec:chernconst}, we explain why the size-independent inversion eigenvalue is forbidden in the two-dimensional HK-type construction due to the Chern-number constraint.
}

\section{Ground state inversion eigenvalue}\label{sec:gs inv ev}

\revise{
This section provides the reference results needed to identify the nonlocal crystalline phase constructed in Sec.~\ref{sec:HK}.
In Sec.~\ref{subsec:setup}, we first specify the setup together with relevant symmetries, and then we introduce the ground-state inversion eigenvalue $I(L)$ used as a diagnostic.
In Sec.~\ref{subsec:0Dview}, we show that $I(L)=-1$ is allowed by symmetry representation theory from the zero-dimensional viewpoint.
In Sec.~\ref{subsec:band}, we show that a noninteracting state must have the trivial inversion eigenvalue $I(L)=+1$.
Finally, in Sec.~\ref{subsec:interacting local}, for local interacting one-dimensional crystalline FSPTs, the real-space construction restricts the inversion eigenvalue to the size-dependent form $I(L)=(\pm1)^L$.
These results identify the size-independent value $I(L)=-1$ as a symmetry-allowed but locality-forbidden quantum number, which will be realized by the
HK-type nonlocal interaction in Sec.~\ref{sec:HK}.
}

\begin{figure}[t]
    \centering
    \includegraphics[width=1\linewidth]{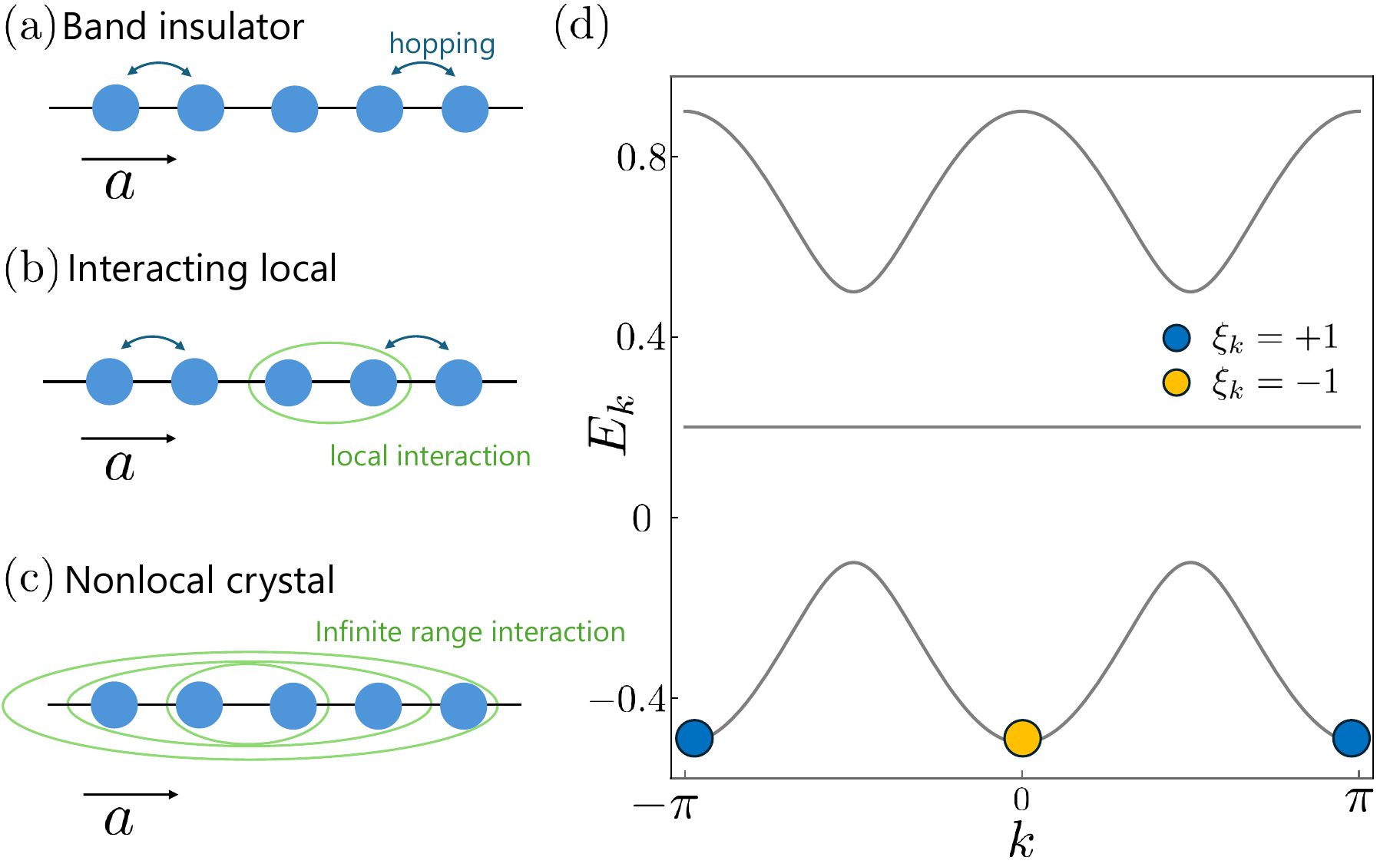}
    \caption{Schematic illustration of hopping and interaction terms in (a) noninteracting phases, namely band insulators, (b) interacting local phases, and (c) interacting nonlocal crystals. Systems invariant under translations by a lattice vector $a$ are considered.
    (d) ``Band structure'' of the nonlocal crystal model in Eq.~\eqref{eq:HK1d-hk}, showing the many-body energies $E_k$ of $\hat{H}_k$ as a function of momentum $k$ in the half-filling sector (two particles for each momentum), at $V_1= 0.7, V_2= 0.3, \revise{\mu=0.1}$. Blue (yellow) circles indicate $+1$ ($-1$) inversion eigenvalue of the ground state, which is defined at $k=\pi$ ($k=0$). 
    }
    \label{fig:HK-band}
\end{figure}

\subsection{Setup and diagnostic}\label{subsec:setup}

\revise{We consider fermions arranged on a finite ring of
$L$ unit cells, equipped with the translation group $\mathbb Z_L$.
With short-range Hamiltonians, this is the usual one-dimensional lattice
setting. 
When locality is abandoned, the systems can instead be viewed as a zero-dimensional quantum system, where translation acts as an internal $\mathbb Z_L$ symmetry at fixed $L$.}

\revise{
We further impose $U(1)$ charge conservation and spinful time-reversal symmetry,
corresponding to symmetry class AII in the ten-fold way
classification~\cite{Ryu-NJP-10,Kitaev-09}, together with translation $\hat{\mathcal{R}}$
and inversion symmetry $\hat{\mathcal I}$. 
For a unique gapped ground state $\ket{\Psi(L)}$, we define
\begin{equation}
    \hat{\mathcal{I}} \ket{\Psi(L)} = I(L) \ket{\Psi(L)},
\end{equation}
where $ \ket{\Psi(L)}$ is the ground state for a chain of $L$ unit cells under periodic boundary conditions, and $I(L)$ is the $L$-dependent inversion eigenvalue of $\ket{\Psi(L)}$ under the action of inversion symmetry.
We use \(I(L)\) to compare the values allowed by fixed-\(L\)
symmetry representation theory with those realizable in noninteracting
band insulators and local interacting crystalline FSPTs.
}



\begin{table}[t]
    \centering
      \caption{Summary of results listing the ground state inversion eigenvalue $I(L)$ and polarization $P$ for (i) noninteracting, (ii) interacting local, and (iii) interacting nonlocal cases in the presence of spinful TRS, charge conservation, and inversion symmetry in 1D.}
    \begin{tabular}{l||c|c}
     \hline
         & $I(L)$ & $P$  \\ \hline
       (i) Noninteracting (local or nonlocal) & $+1$ & 0   \\
       (ii) Interacting local & $(\pm 1)^L$& 0  \\
       (iii) Interacting nonlocal crystal & $(\pm 1)^L, \ -1$ & $1/2$ \\
         \hline
    \end{tabular}
    \label{tab:summary-I-Q}
\end{table}

\subsection{Fixed-$L$ zero-dimensional viewpoint}
\label{subsec:0Dview}
\revise{
We discuss which inversion eigenvalues are allowed by symmetry
alone, without imposing locality. At fixed $L$, we 
consider a translation operator acting as a unitary internal $\mathbb Z_L$
symmetry on the Hilbert space.
}

\revise{
The corresponding fixed-$L$ symmetry can be written as
\begin{equation}
    G_{f,L}
    =
    G_f^{\rm AII}
    \times
    \left(\mathbb Z_L \rtimes \mathbb Z_2^{\mathcal I}\right),
    \quad
    G_f^{\rm AII}
    =
    \frac{U(1)^F \rtimes \mathbb Z_4^{FT}}{\mathbb Z_2^F}.
\end{equation}
Here, \(G_f^{\rm AII}\) is the internal fermionic symmetry of class AII~\cite{Soldini-PRR-2024}:
\(U(1)^F\) is generated by
\(\hat U(\theta)=e^{i\theta\hat N}\), and
\(\mathbb Z_4^{FT}\) encodes spinful time reversal
\(\hat{\mathcal T}^2=(-1)^{\hat N}\), with the common fermion parity
\(\mathbb Z_2^F\) identified in the quotient. The crystalline part is
generated by translation \(\hat{\mathcal{R}}\) and inversion \(\hat{\mathcal I}\).
The nontrivial actions are
\begin{equation}
    \hat{\mathcal I}\hat{\mathcal{R}}\hat{\mathcal I}^{-1}
    =
    \hat{\mathcal{R}}^{-1},
    \qquad
    \hat{\mathcal T}\hat U(\theta)\hat{\mathcal T}^{-1}
    =
    \hat U(-\theta),
\end{equation}
with all remaining actions taken to be trivial. The second relation
follows from the antiunitarity of \(\hat{\mathcal T}\).
}

\revise{
For a unique gapped ground state $\ket{\Psi(L)}$, we write its symmetry
quantum numbers as
\begin{equation}
\begin{aligned}
    \hat{U}(\theta)\ket{\Psi(L)} &= e^{iN\theta}\ket{\Psi(L)},\\
    \hat{\mathcal{R}}\ket{\Psi(L)} &= e^{2\pi i p/L}\ket{\Psi(L)},\\
    \hat{\mathcal I} \ket{\Psi(L)} &= \eta_I\ket{\Psi(L)} .
\end{aligned}
\end{equation}
Here, $N$ is the total charge eigenvalue, $p\in\mathbb Z_L$ is the total crystal momentum, and $\eta_I=\pm1$ is the inversion eigenvalue. Since
$\hat{\mathcal I}$ maps momentum $p$ to $-p$, a nondegenerate
inversion-symmetric ground state must satisfy
\begin{equation}
    2p=0 \mod L .
\end{equation}
Thus $p=0$ is always allowed, while $p=L/2$ is also possible for even
$L$.
}

\revise{
In the following, we focus on the sector relevant to the ground states
considered in this work: fixed even charge and total momentum $p=0$.
The even-charge condition is compatible with spinful time-reversal
symmetry, for which $\hat{\mathcal T}^2=(-1)^{\hat{N}}$, and avoids a Kramers degeneracy of the ground state. 
Within this fixed sector, the symmetry
algebra imposes no further constraint on $\eta_I$. Therefore both
\begin{equation}
    \eta_I=+1,\qquad \eta_I=-1
\end{equation}
are allowed by fixed-$L$ symmetry representation theory. In this
sector, the ground-state inversion eigenvalue introduced above is simply
$I(L)=\eta_I$.
}

\subsection{Band insulators}\label{subsec:band}
\revise{Next, we consider the noninteracting case [Fig.~\ref{fig:HK-band} (a)].}
In systems with spinful TRS, any gapped ground state forms a single Slater determinant of Kramers pairs, which transform trivially under spatial symmetries since the two electrons in each pair carry complex-conjugate eigenvalues~\cite{Yao-10-PRL}.
In particular, when acting with the inversion operator on the ground state, the inversion eigenvalue is
\begin{equation}
    I(L) = +1,
\end{equation}
irrespective of the details of the Hamiltonian, and the system size $L$.
Thus the inversion eigenvalue is always trivial for band insulators.

\subsection{Interacting local phases}
\label{subsec:interacting local}
We now introduce interactions while keeping the Hamiltonian local [Fig.~\ref{fig:HK-band} (b)]. In 1D, there are no nontrivial FSPTs with the symmetries of class AII~\cite{Freed2021,Aksoy2023,note-1D-topo}. 
Including inversion and translation symmetries, 
1D crystalline FSPTs can be classified by the real-space construction 
~\cite{Song-PRX-17,Huang-PRB-17,Zhang-PRB-20,zhang-arXiv-22,Zhang-PRR-22}. Specifically, such classification is given by enumerating all the distinct reference states compatible with the global and crystalline symmetries, which in this case can be obtained by considering different zero-dimensional (0D) block ``decorations'' with states ``charged'' under inversion, meaning that the states transform with eigenvalue $\pm 1$ under inversion. In fact, there are no topologically nontrivial 1D phases in class AII, and inversion leaves only isolated points of the lattice invariant, therefore it can only protect 0D FSPTs located at the inversion invariant points. There are two locations in the unit cell of the lattice that can be decorated by 0D FSPTs: the center and the edge of the unit cell, corresponding to the $1a$ and $1b$ Wyckoff positions, respectively. By ``decorating 0D blocks'' we specifically mean stacking states localized at points in the lattice corresponding to either the $1a$ or $1b$ Wyckoff positions.
The reference state of the crystalline FSPTs obtained by decorating 0D blocks can be written as
\begin{equation}\label{eq:cFSPTs in 1D}
    \ket{\Psi(L)} = \bigotimes_{i=0}^{L-1} \ket{\psi_{i, 1a}}\ket{\psi_{i, 1b}}.
\end{equation}
Here, $\ket{\psi_{i, 1a/1b}}$ indicates an electronic state localized at either the $1a$ or $1b$ Wyckoff positions of the $i$-th unit cell, on which inversion acts as
\begin{equation}\label{eq:0D decorations under I}
    \hat{\mathcal{I}} \ket{\psi_{i, 1a}} = \pm \ket{\psi_{-i, 1a}}, \quad \hat{\mathcal{I}} \ket{\psi_{i, 1b}} = \pm \ket{\psi_{-i, 1b}} . 
\end{equation}
Any of the inversion eigenvalues in Eq.~\eqref{eq:0D decorations under I} can be explicitly realized in a class of reference states known as Mott atomic limits (MALs) with the symmetries of class AII~\cite{Soldini-PRB-2023,Soldini-PRR-2024}, which take the form $\ket{\mathrm{MAL}}=\prod_{i} \hat{O}^{\dagger}_{i} \ket{\mathrm{vac}}$, where $\ket{\rm vac}$ denotes the vacuum state, and $\hat{O}^{\dagger}_i$ creates $n$-electrons localized in the $i$-th unit cell, and have the advantage that they can be generically obtained as a gapped ground state of a local Hubbard Hamiltonian~\cite{Soldini-PRB-2023,Herzog-Ncom-24}. While 1D band insulators form a subset of MALs in class AII, MALs can encode interaction-enabled states that cannot be connected to any noninteracting states~\cite{Soldini-PRB-2023,Yao-10-PRL}.

As an example, consider two spinful orbitals, $s$ and $p$, with even and odd parity under inversion, at the $1a$ Wyckoff position.
The local state
\begin{equation}\label{eq:MAL at 1a}
    \ket{\psi_{i, 1a}} = \frac{1}{\sqrt{2}}(\hat{c}^{\dagger}_{i, 1a, s, \uparrow} \hat{c}^{\dagger}_{i, 1a, p, \downarrow} - \hat{c}^{\dagger}_{i, 1a, s, \downarrow} \hat{c}^{\dagger}_{i, 1a, p, \uparrow}) \ket{\rm vac}
\end{equation}
respects TRS and transforms as $\hat{\mathcal{I}}\ket{\psi_{i, 1a}} = - \ket{\psi_{-i, 1a}}$ under the inversion.
Here, $\hat{c}^{\dagger}_{i, 1a, \alpha, \sigma}$ ($\hat{c}_{i, 1a, \alpha, \sigma}$) creates (annihilates) an electron at the $1a$ Wyckoff position of the $i$-th unit cell, with orbital $\alpha\in\{s, p\}$ and spin $\sigma\in\{\uparrow, \downarrow\}$.
An analogous construction can be repeated for the $1b$ Wyckoff position. 
From Eqs.~\eqref{eq:cFSPTs in 1D} and~\eqref{eq:0D decorations under I}, we deduce that the ground state of a crystalline FSPT in class AII with inversion symmetry transforms under the action of inversion symmetry as
\begin{equation}
    I(L) = (\pm 1)^L
\end{equation}
where $(-1)^L$ holds if $\ket{\psi_{1a}}$ and $\ket{\psi_{1b}}$ transform with the opposite sign under inversion, and $(+1)^L$ holds otherwise.

\revise{
Thus, the size-independent pattern $I(L)=-1$ is allowed by
fixed-$L$ nonlocal (effectively zero-dimensional) symmetry representation theory, but is
excluded by the classification of local one-dimensional crystalline
FSPTs. In the following, we show that an HK-type nonlocal interaction
realizes this symmetry-allowed but locality-forbidden inversion eigenvalue as a unique gapped ground state.
}


\section{Nonlocal crystalline model}
\label{sec:HK}

\subsection{Model}
\revise{In this section, we construct a minimal nonlocal model whose ground state realizes the size-independent inversion eigenvalue \(I(L)=-1\)  [Fig.~\ref{fig:HK-band} (c)].}
We employ an exactly solvable HK-type interaction~\cite{HK-92-JPSJ}, which is
diagonal in momentum space and therefore arbitrarily long-ranged in real
space. HK-type models have recently been used to study interacting
topological phases~\cite{Mai-PRR-23,Zhao-PRL-23,Mai-Ncom-23,Manning-PRB-23,Krystian-PRB-23,Mai-PRB-24,Qi-arxiv-24,Hooley-arxiv-24,Skolimowski-PRB-25,Sinha-PRB-25},
where the topological structure usually originates from local one-body
terms. 
\revise{In contrast, our construction uses the HK-type interaction to realize the symmetry-allowed but locality-forbidden inversion eigenvalue.}

Following Hatsugai and Kohmoto~\cite{HK-92-JPSJ}, we consider a
momentum-space diagonal Hamiltonian \(\hat H=\sum_k \hat H_k\) for a 1D
spinful system with two orbitals, \(s\) and \(p\), located at the \(1a\)
Wyckoff position. The two orbitals are even and odd under inversion,
respectively [Fig.~\ref{fig:HK-band}(c)]. We denote the creation and
annihilation operators by \(\hat c^\dagger_{k,\alpha,\sigma}\) and
\(\hat c_{k,\alpha,\sigma}\), where \(\alpha\in\{s,p\}\) and
\(\sigma\in\{\uparrow,\downarrow\}\). The Hamiltonian reads
\begin{align}\label{eq:HK1d-hk}
     &\hat{H}_k = V_1\,\cos{k}\left[ \hat{C}^\dag_{2}(k) \hat{C}_{2}(k)- \hat{C}_1^{\dagger}(k) \hat{C}_1(k)\right]   \\
    +&\ii V_2\, \sin{k} \left[ \hat{C}_2^{\dagger}(k) \hat{C}_{1}(k) -\hat{C}_1^{\dagger}(k) \hat{C}_{2}(k)\right]    \revise{+ \mu \sum_{\substack{\alpha=s,p \\ \sigma=\uparrow, \downarrow}} \hat{c}_{k,\alpha,\sigma}^\dag \hat{c}_{k,\alpha,\sigma} } \nonumber,
\end{align}
with real parameters \(V_1,V_2>0\), and we have introduced
\begin{align}
     \hat{C}^{\dagger}_{1} (k) &= \frac{1}{\sqrt{2}} ( 
    \hat{c}^\dag_{k, s, \uparrow }\hat{c}^\dag_{k, p, \downarrow } -
    \hat{c}^\dag_{k, s, \downarrow }\hat{c}^\dag_{k, p, \uparrow } ), \\
    \hat{C}^{\dagger}_2(k) &= \hat{c}^\dag_{k, s, \uparrow }\hat{c}^\dag_{k, s, \downarrow }, 
\end{align}
which are TRS and transform under inversion as
\begin{equation}
    \hat{\mathcal{I}} \hat{C}^{\dagger}_{1}(k) \hat{\mathcal{I}}^{-1} = -\hat{C}^{\dagger}_{1}(-k), ~~
    \hat{\mathcal{I}} \hat{C}^{\dagger}_2(k) \hat{\mathcal{I}}^{-1} = \hat{C}^{\dagger}_2(-k).
\end{equation}
\revise{The chemical potential term ($0 < 2\mu < V_1, V_2$) ensures that the ground state exists in the half-filling sector.}
The Hamiltonian in Eq.~\eqref{eq:HK1d-hk} preserves TRS and inversion symmetry 
\begin{equation}
  [\hat{H},\hat{\mathcal{T}}] = [\hat{H}, \hat{\mathcal{I}}]=0.
\end{equation}

\subsection{Ground state and inversion eigenvalue}

Here we obtain the exact ground state in the half-filling sector and compute the ground state inversion eigenvalue.
Since each momentum $k$ has four internal states originating from the two spinful orbitals and is occupied by two electrons, the Hilbert space dimension is $\dim{\mathcal{H}_k} = 6$.
The Hamiltonian $(\hat{H}_k )$ in Eq.~\eqref{eq:HK1d-hk} thus acts on a six-dimensional Hilbert space and takes the matrix form
\begin{align}\label{aeq:HK0-matrix}
\hat{H}_k = 
 \left(
\begin{array}{cc|cccc} 
     -V_1 \cos{k} \revise{+ 2\mu} &  -\ii V_2 \sin{k} &  0 & 0 & 0& 0 \\ 
   \ii V_2 \sin{k} & V_1 \cos{k} \revise{+2\mu} & 0 & 0 & 0 & 0 \\  \hline 
0 &0 & \revise{2\mu} & 0 & 0 & 0 \\
0 & 0& 0& \revise{2\mu} &0 &0 \\
0 & 0& 0& 0 &\revise{2\mu} &0 \\
0 & 0& 0& 0 &0 &\revise{2\mu} 
\end{array} 
\right)
\end{align}
with the six basis states obtained by applying the following operators to the vacuum state $\ket{\rm vac}$
\begin{align}\label{aeq:basis}
     \Big\{\hat{C}^\dag_{1}(k),~\hat{C}^\dag_{2}(k),
      \frac{1}{\sqrt{2}} ( 
    \hat{c}^\dag_{k, s, \uparrow }\hat{c}^\dag_{k, p, \downarrow } +
    \hat{c}^\dag_{k, s, \downarrow }\hat{c}^\dag_{k, p, \uparrow } ) ,\nonumber \\
    ~\hat{c}^\dag_{k, p, \uparrow }\hat{c}^\dag_{k, p, \downarrow },~\hat{c}^\dag_{k, s, \uparrow }\hat{c}^\dag_{k, p, \uparrow },~\hat{c}^\dag_{k, s, \downarrow }\hat{c}^\dag_{k, p, \downarrow }
     \Big\}.
\end{align}
By diagonalizing the matrix in Eq.~\eqref{aeq:HK0-matrix}, we obtain six eigenvalues
\begin{equation}
    E_k =  \revise{2\mu} \pm \sqrt{(V_1 \cos{k})^2 + (V_2 \sin{k})^2}, ~\revise{2\mu}.
\end{equation}
Note that $E_k=2\mu$ is fourfold degenerate.
The ground-state energy is given by $E_k=\revise{2\mu}  -\sqrt{(V_1 \cos{k})^2 + (V_2 \sin{k})^2}$ and the corresponding eigenstate takes the form 
\begin{align}\label{aeq:gs-HK}
    \ket{\psi(k)} &= \left( 
    \psi_{1}(k) \hat{C}_{1}^\dag (k) + \psi_{2}(k) \hat{C}_{2}^\dag (k) 
    \right) \ket{{\rm vac}}
\end{align}
with 
\begin{align}
     \psi_1(k) = \ii \cos{\frac{\varphi(k)}{2}}, \quad
   \psi_2(k) = \sin{\frac{\varphi(k)}{2}}.
\end{align}
Here we have introduced 
\begin{equation}
   \varphi(k) = \tan^{-1} \Big( \frac{V_2}{V_1} \tan{k} \Big). 
\end{equation}

Thus, the total ground state of the full Hamiltonian $\ket{\Psi} = \mathop{\bigotimes}_{k \in \rm{BZ}} \ket{\psi(k)}$ is given by
\begin{align}\label{eq:total-gs-HK}
     \ket{\Psi} = \prod_{k \in {\rm BZ}}\left( 
    \psi_{1}(k) \hat{C}_{1}^\dag (k) + \psi_{2}(k) \hat{C}_{2}^\dag (k) 
    \right) \ket{{\rm vac}}.
\end{align}

We now compute the inversion eigenvalue of this ground state. Since
inversion maps \(k\) to \(-k\), the contributions from generic momenta
\(k\neq0,\pi\) cancel pairwise in the product state. Therefore, the
many-body inversion eigenvalue is determined only by the
inversion-invariant momenta:
\begin{equation}
\label{eq:I-L-HK}
I(L)
=
\begin{cases}
\xi_{k=0}\xi_{k=\pi}, & L \ {\rm even},\\
\xi_{k=0}, & L \ {\rm odd},
\end{cases}
\end{equation}
where \(\hat{\mathcal I}\ket{\psi(k)}=\xi_k\ket{\psi(k)}\) at
\(k=0,\pi\).
From Eq.~\eqref{eq:total-gs-HK}, the ground state at \(k=0\) is
\(\ii\hat C_1^\dagger(0)\ket{\rm vac}\), whereas the ground state at
\(k=\pi\) is \(\hat C_2^\dagger(\pi)\ket{\rm vac}\). Using the inversion
transformation of \(\hat C_1^\dagger\) and \(\hat C_2^\dagger\), we find
\begin{equation}
\xi_{k=0}=-1,
\qquad
\xi_{k=\pi}=+1.
\end{equation}
Thus, for both even and odd \(L\), the many-body ground state has
\begin{equation}
I(L)=-1.
\end{equation}
\revise{
This realizes the size-independent inversion eigenvalue that is allowed
by fixed-size symmetry representation theory but forbidden in
noninteracting band insulators and local interacting one-dimensional
crystalline FSPTs.
}

\subsection{Gap closing for the local limit}
\label{subsec:gap}
\begin{figure}[t]
\centering
\includegraphics[width=0.8\linewidth]{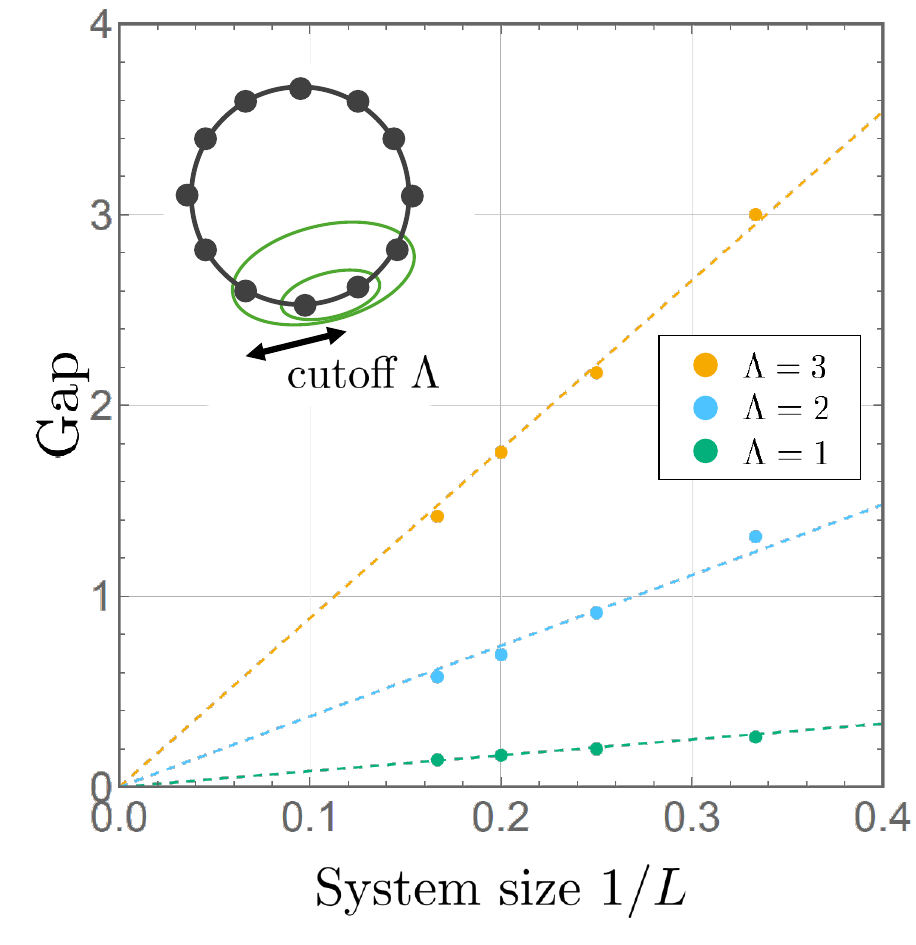} 
\caption{System-size dependence of the energy gap between the ground state and first excited states in the half-filling sector, as obtained from the exact diagonalization ($V_1=V_2=1.0, \mu=0.1$). The interaction range cutoff $\Lambda$ is chosen as $\Lambda=3$ (orange), $\Lambda=2$ (blue), and $\Lambda = 1$ (green).
} \label{fig: gapL}
\end{figure}

Finally, we examine what happens when locality is imposed on the
HK-type interaction. We restrict the interaction range in real space to a
finite cutoff \(\Lambda\) and study the resulting finite-size gap
(Fig.~\ref{fig: gapL}).
A subtle but rarely discussed issue arises when imposing locality on HK-type interactions: fixing the interaction range $\Lambda$ leads to a ground-state energy scaling $E_{\mathrm{GS}} \sim \mathcal{O}(L^0)$, violating the expected energy extensivity of local Hamiltonians, which would require $E_{\mathrm{GS}} \sim O(L^1)$ in 1D. To restore extensivity, we multiply the HK-type interaction by the system size $L$ (see Appendix~\ref{asec:hkreal}).
Figure~\ref{fig: gapL} shows the system-size dependence of the energy gap between the ground state and the first excited state at half-filling for interaction range cutoffs $\Lambda=1, 2, 3$.
In all cases, the energy gap approaches zero for $L \rightarrow \infty$, indicating that the ground state is gapless. 
This supports that
the gapped realization of the size-independent inversion eigenvalue
\(I(L)=-1\) relies on the fully nonlocal HK-type interaction.

\section{Resta's formula under periodic boundary conditions}
\label{sec:berry}

\revise{
In Sec.~\ref{sec:HK}, we constructed a unique gapped ground state with
the symmetry-allowed but locality-forbidden inversion eigenvalue
\(I(L)=-1\). We now ask what additional ground-state information can be
extracted from this state under periodic boundary conditions.
}

\revise{
For fully nonlocal Hamiltonians, neither open-boundary diagnostics nor
the coupling to background gauge fields is canonical. Open boundaries
are ambiguous because the interaction range scales with the system size,
and a Peierls-type prescription for introducing gauge fields is not
uniquely defined for arbitrary-range many-body interactions.
}

\revise{
We therefore focus on Resta's formula, which defines a periodic boundary conditions analogue of the many-body position directly from the finite-size ground-state wave function~\cite{Resta-PRL-98}:
\begin{equation} P = \frac{1}{2\pi}\mathrm{Im}\log \bra{\Psi}\hat z\ket{\Psi}, \qquad \hat z = \exp\left[ \frac{2\pi i}{L}\sum_x x\hat n_x \right], \label{eq:Resta} 
\end{equation}
where \(\hat n_x\) denotes the local particle number operator. The
quantity \(P\) is defined modulo one. The coordinate \(x\) includes the
intra-unit-cell position of the corresponding orbital.
Intuitively, $P$ captures the center of mass of the ground-state electron distribution under periodic boundary conditions.
In ordinary local insulators, it is identified with the electric
polarization.}

\imp{We stress that, although Eq.~\eqref{eq:Resta} is well defined as a
geometric quantity of the finite-size ground state, there is
no general basis for identifying this quantity with a physical
electric polarization in the fully nonlocal setting considered
here. Likewise, its adiabatic variation cannot in general be
identified with transported charge. The conventional identification relies on locality in real space, for which the coupling to an external electromagnetic field and the associated current response are well defined. For fully nonlocal HK-type interactions, however, it is not obvious how the interaction should be coupled to an inserted flux. Additionally, the current inferred from the continuity
equation is singular in the long-wavelength limit
~\cite{Guerci}. In the present work, we therefore use
Eq.~(27) only as a geometric quantity of the finite-size
ground state under periodic boundary conditions for the
explicitly constructed HK-type solvable state.}

\revise{
Before evaluating the HK-type ground state, we first consider the two local reference classes discussed in Sec.~\ref{sec:gs inv ev}.
In noninteracting band
insulators in class AII, occupied states form Kramers pairs. Inversion
symmetry quantizes the Berry phase of each member of a Kramers pair to
\(0\) or \(1/2\), while TRS makes the two partners carry the same Berry
phase. Thus each Kramers pair contributes either \(0\) or \(1\) to
\(P\), and the total Resta quantity is (Table~\ref{tab:summary-I-Q})
\begin{equation}
    P=0 \quad \mathrm{mod}\ 1 .
\end{equation}
For local interacting one-dimensional crystalline FSPTs, the real-space construction
reduces the calculation to MALs. A symmetric local
block compatible with spinful TRS carries an even charge; otherwise an
odd-charge block would form a Kramers doublet. Therefore, both the
\(1a\) and \(1b\) Wyckoff-position blocks contribute an integer to
Eq.~\eqref{eq:Resta}, and the corresponding local interacting phases
also have (Table~\ref{tab:summary-I-Q})
\begin{equation}
    P=0 \quad \mathrm{mod}\ 1 .
\end{equation}
}

\revise{
We now evaluate the same quantity for the HK-type ground state in
Eq.~\eqref{eq:total-gs-HK}. Since the twist operator shifts the momentum
of each creation operator by \(\Delta k=2\pi/L\), the pair creation
operators obey
\begin{equation}
    \hat z\hat C^\dagger_j(k)\hat z^{-1}
    =
    \hat C^\dagger_j(k+\Delta k).
\end{equation}
Therefore, the expectation value of \(\hat z\) factorizes into a
discrete Berry phase of the two-component wave function
\(\ket{\psi(k)}\):
\begin{align}
    \bra{\Psi}\hat z\ket{\Psi}
    &=
    \prod_{k\in{\rm BZ}}
    \braket{\psi(k+\Delta k)}{\psi(k)}
    \label{eq:Resta-product}
\end{align}
Taking the thermodynamic limit, Eq.~\eqref{eq:Resta-product} gives
\begin{equation}
    P
    =
    \frac{1}{2\pi}
    \oint_{\rm BZ} dk\,
    \langle \psi(k)|\ii \partial_k|\psi(k) \rangle
    \label{eq:Resta-Berry}
\end{equation}
}

\revise{
The Berry phase in Eq.~\eqref{eq:Resta-Berry} is fixed by inversion
eigenvalues at the inversion-invariant momenta. Indeed, inversion
symmetry implies $\ket{\psi(-k)} = e^{i\theta(k)} \hat{\mathcal I}\ket{\psi(k)}$,
and hence
\begin{equation}
    P
    =
    \frac{i}{2\pi}
    \log
    \left[
    \frac{\xi_{k=\pi}}{\xi_{k=0}}
    \right]
    \quad
    \mathrm{mod}\ 1,
    \label{eq:Resta-inversion}
\end{equation}
where
\(\hat{\mathcal I}\ket{\psi(k_0)}
=
\xi_{k_0}\ket{\psi(k_0)}\) for \(k_0=0,\pi\).
For the HK-type ground state in Eq.~\eqref{eq:total-gs-HK}, we found
\(\xi_{k=0}=-1\) and \(\xi_{k=\pi}=+1\). Thus we have (Table~\ref{tab:summary-I-Q})
\begin{equation}
    P=\frac{1}{2}.
\end{equation}
}

\revise{
Therefore, under periodic boundary conditions, the HK-type ground state is distinguished not only by the locality-forbidden inversion eigenvalue \(I(L)=-1\), but also by a quantized many-body center-of-mass position. Resta's formula assigns this state the nontrivial value \(P=1/2\), indicating that the many-body center of mass is pinned to the nontrivial inversion-symmetric position.
}

\section{Generalization}\label{sec:generalization}
\begin{table}[t]
    \centering
     \caption{Allowed inversion eigenvalue patterns for BIs, MALs, and nonlocal crystalline phases.
    In the entries, $\checkmark$ indicates that there exists a model realizing the specified inversion eigenvalue, and $-$ indicates the absence of such a model. 
    }
    \begin{tabular}{c|ccc|cccc}
    \hline \hline
         & \multicolumn{3}{c|}{1D} & \multicolumn{3}{c}{2D} \\
        $I(L)$ & $+1$ & $(-1)^L$ & $-1$ & $+1$ & $(-1)^{L_x L_y}$ & $(-1)^{L_{x/y}}$ & $-1$\\
        \hline
        BI & \checkmark & -- & -- & \checkmark & -- & -- & --  \\
        MAL & \checkmark & \checkmark & -- & \checkmark & \checkmark & -- & -- \\
        HK  & \checkmark & \checkmark & \checkmark & \checkmark & \checkmark & \checkmark & -- \\
        \hline \hline
    \end{tabular}
    \label{tab:inversion-eigenvalues}
\end{table}
While we have focused on inversion symmetry in 1D, our construction naturally generalizes to higher spatial dimensions. 
In Table~\ref{tab:inversion-eigenvalues}, we list the possible inversion eigenvalues $I(L)$ for systems in 1D and $I(L_x,L_y)$ in 2D.
\revise{
The 1D case has been discussed in detail in this paper, and 
the central result is that \(I(L)=-1\) is excluded in the noninteracting and local FSPTs settings considered above, but is realized by the HK-type nonlocal model.
}

In 2D, we find that $I(L_x,L_y) = (-1)^{L_x L_y}$ can be obtained with the real space construction of crystalline FSPTs, i.\,e., by tiling the system with 0D localized wavefunctions with odd inversion parity. Similarly, stacking 1D chains of the 1D nonlocal crystalline phase of Eq.~\eqref{eq:HK1d-hk} in real space along one direction in a weak topological fashion leads to $I(L_x,L_y) = (-1)^{L_x}$ or $(-1)^{L_y}$.
Interestingly, in contrast to the 1D case, we find that $I(L_x,L_y) = -1$ cannot be realized, even with the HK-type interaction, due to constraints imposed by the Chern number (Appendix~\ref {asec:chernconst}). 
Extensions to three or higher dimensions and other point group symmetries can also be cataloged within our framework.

\section{Discussion}\label{sec:discussion}

\revise{
In this paper, we have shown that HK-type nonlocal interactions provide an explicit route to realizing ground-state symmetry sectors that are allowed by fixed-size symmetry representation theory but forbidden by locality in local one-dimensional crystalline FSPTs. In particular, we constructed an exactly solvable HK-type model whose unique gapped translation-invariant ground state has the system-size-independent inversion eigenvalue $I(L)=-1$, which is inaccessible in both noninteracting band insulators and local interacting one-dimensional crystalline FSPTs with the same symmetries. We also evaluated Resta's formula for the constructed HK-type ground state under periodic boundary conditions and found that the corresponding finite-size ground-state quantity takes the quantized value $P=1/2$. We further discussed how the comparison between fixed-size symmetry representation theory, noninteracting band theory, and local crystalline FSPTs extends to higher dimensions.
}

\revise{
Long-range interactions realized in trapped ions, ultracold atoms, and Rydberg arrays~\cite{Defenu-RMP-23,Blatt-NPhys-12,Maucher-PRL-11,Saffman-RMP-10} may enable the experimental realization of nonlocal crystalline models.
Synthetic dimensions offer another promising route, where tunable components in 0D systems emulate HK-type interactions and have already supported both noninteracting~\cite{Lohse-Nat-18} and many-body models~\cite{An-PRL-21}.
}

\medskip
\section*{Acknowledgments}
We thank \"{O}mer M.~Aksoy, Frank Schindler, and Ken Shiozaki for useful discussions.
S.H.~is supported by JSPS Research Fellow No.~24KJ1445, JSPS Overseas Challenge Program for Young Researchers, and the MEXT WISE program.
S.H.~and T.Y.~are grateful for the support and hospitality of the Pauli Center for Theoretical Studies.
T.Y. is supported by JSPS KAKENHI Grant Nos. JP21K13850, and JP23KK0247, JSPS
Bilateral Program No. JPJSBP120249925 and the Grant from Yamada Science Foundation.
T.N. and M.S. acknowledge support from the Swiss National Science Foundation through a Consolidator Grant (iTQC, TMCG-2213805).

\appendix


\section{Locality of the HK-type interaction in real space}
    \label{asec:hkreal}
In this section, we provide a detailed procedure to impose locality on a Hamiltonian with an HK-type interaction in real space.
A generic HK-type interaction term contains operators of the form $\sum_k \hat{c}^\dag_{k, \alpha} \hat{c}^\dag_{k, \beta} \hat{c}_{k, \gamma} \hat{c}_{k, \delta}$.
From the Fourier transformation of the creation and annihilation operators
\begin{align}
    \hat{c}^\dag_{k, \alpha} = \frac{1}{\sqrt{L}} \sum_{x} e^{\ii k x} \hat{c}^\dag_{x, \alpha}, \quad \hat{c}_{k, \alpha} = \frac{1}{\sqrt{L}} \sum_{x} e^{-\ii k x} \hat{c}_{x, \alpha},
\end{align}
we have 
\begin{widetext}
\begin{align}\label{aeq:FT of the HK term}
    \sum_k \hat{c}^\dag_{k, \alpha} \hat{c}^\dag_{k, \beta} \hat{c}_{k, \gamma} \hat{c}_{k, \delta}
    &=\frac{1}{L^2}
    \sum_k \sum_{x_1, \cdots, x_4} e^{\ii k(x_1 + x_2 -x_3 -x_4)}
    \hat{c}^\dag_{x_1, \alpha} \hat{c}^\dag_{x_2,\beta}
    \hat{c}_{ x_3,\gamma} \hat{c}_{ x_4,\delta} \nonumber \\
    &=\frac{1}{L}\sum_{x_1, \cdots, x_4}\delta(x_1 +x_2 -x_3 -x_4)  \hat{c}^\dag_{x_1, \alpha} \hat{c}^\dag_{x_2,\beta}
    \hat{c}_{ x_3,\gamma} \hat{c}_{ x_4,\delta}.
\end{align}
\end{widetext}
The delta function in the last line implies that the term in Eq.~\eqref{aeq:FT of the HK term} preserves the center of mass of the electronic states that are created and annihilated. Therefore, we can rewrite the terms as
\begin{widetext}
\begin{align}\label{aeq:rangelambda}
     \sum_k \hat{c}^\dag_{k, \alpha} \hat{c}^\dag_{k, \beta} \hat{c}_{k, \gamma} \hat{c}_{k, \delta} 
     =  \frac{1}{8L} \sum_{R=0}^{2L-1} \sideset{}{'}\sum_{r_1, r_2} \left[
     \hat{c}^\dag_{ \frac{R+r_1}{2}, \alpha} \hat{c}^\dag_{\frac{R-r_1}{2}, \beta}
    \hat{c}_{\frac{R+r_2}{2}, \gamma } \hat{c}_{ \frac{R-r_2}{2}, \delta}
    +
     \hat{c}^\dag_{\frac{R-r_1}{2}, \alpha } \hat{c}^\dag_{\frac{R+r_1}{2}, \beta }
    \hat{c}_{\frac{R-r_2}{2}, \gamma } \hat{c}_{ \frac{R+r_2}{2},\delta} \right. \nonumber \\
    + \left.
     \hat{c}^\dag_{\frac{R+r_1}{2}, \alpha } \hat{c}^\dag_{ \frac{R-r_1}{2}, \beta}
    \hat{c}_{\frac{R-r_2}{2}, \gamma } \hat{c}_{\frac{R+r_2}{2}, \delta } +
     \hat{c}^\dag_{\frac{R-r_1}{2},\alpha } \hat{c}^\dag_{ \frac{R+r_1}{2}, \beta}
    \hat{c}_{\frac{R+r_2}{2},\gamma } \hat{c}_{ \frac{R-r_2}{2},\delta} \right].
\end{align}
\end{widetext}
where we have introduced the center of mass coordinate $R/2$ of a pair of created or annihilated electrons, with $R$ ranging from $0$ to $2L-1$.
The summation over the variables $r_1, r_2$ is restricted to the following ranges, depending on the parity of $R$
\begin{align}
    &{\rm{If~R~is~odd:}}~~ r_{1}, r_2 =1,3,\cdots,2\Lambda-1 \\
    &{\rm{If~R~is~even:}}~~ r_1, r_2 =0,2,\cdots,2\Lambda-2.
\end{align}
The value $\Lambda$ specifies the cutoff for the interaction range, and $\Lambda=L$ corresponds to the original HK-type interaction. 

The factor $1/8$ in Eq.~\eqref{aeq:rangelambda} originates from the following considerations:
In Eq.~\eqref{aeq:rangelambda}, since we need to maintain the center of mass at $R/2$, we must sum over all allowed combinations.
If only the first term within the brackets $[\star]$ is considered, 
it always creates (annihilates) the fermions with internal degrees of freedom $\alpha$ ($\gamma$) at locations greater than $R/2$ and with internal degrees of freedom $\beta$ ($\delta$) at locations smaller than $R/2$.
This incomplete coverage of the configuration space breaks the translation invariance when $\Lambda < L$.
Hence, all four permutations must be summed.
Furthermore, a given configuration appears in two sets of parameters: $\{R, r_1, r_2\}, \{R+L, r_1+L, r_2+L\}$.
As a result, the overall factor becomes $1/4 \times 1/2 = 1/8 $.

As a remark, we point out a subtle issue that arises from the inherent nature of the HK-type interaction when locality is imposed on the Hamiltonian.  
Once locality is imposed (i.e., the interaction range $\Lambda$ is fixed irrespective of the system size $L$), the interaction term in Eq.~\eqref{aeq:rangelambda} remains of order $\mathcal{O}(L^0)$.  
\revise{This is because a prefactor proportional to $1/L$ appears, as shown in Eq.~\eqref{aeq:rangelambda}, when the HK-type interaction is Fourier transformed into real space.}
However, for a Hamiltonian composed of local terms, it is natural to expect that the ground state energy scales extensively with system size, in accordance with the standard assumptions of thermodynamics.  
Note that this deviation from extensivity is not a peculiarity of our specific model, but rather an intrinsic property of HK-type models, although it has rarely been discussed in the literature.  
To address this aspect, we multiply the HK-type interaction term in the Hamiltonian by a factor of $L$ in our numerical simulations to restore extensivity.  
In Fig.~\ref{fig: gapL} of the main text, we follow the procedure described above to construct the real-space representation of the Hamiltonian in Eq.~\eqref{eq:HK1d-hk} with a finite interaction range.

\revise{Additionally, we note that not all nonlocal models suffer from a lack of energy extensivity.
Indeed, as discussed in Sec. I.A of Ref.~\cite{Defenu-RMP-23}, when the exponent $\alpha$ satisfies $\alpha > d$ for interactions of the form 
$V(r) \simeq 1/r^\alpha$ in spatial dimension $d$, the energy extensivity is preserved despite the long-range nature of the interaction.}

\section{Classification of FSPTs by real-space construction}
\label{asec:classification}
For completeness, here we report the FSPT classification of 1D systems in class AII with inversion and lattice translation symmetries, as obtained from the real-space construction~\cite{Song-PRX-17,Huang-PRB-17,Zhang-PRB-20,zhang-arXiv-22,Zhang-PRR-22}. Note that we omit the $2\mathbb{Z}$ term in the classification counting the total number of electrons in the unit cell.

\subsection{Noninteracting case}
For the noninteracting case, the topological classification is $2\mathbb{Z} \times 2\mathbb{Z}$.
Each $2\mathbb{Z}$ factor in the classification corresponds to the difference in the number of single-particle states with inversion eigenvalues $\pm 1$ localized at the $1a$ or the $1b$ Wyckoff positions, respectively. 

\subsection{Interacting case}
For the interacting case, the classification is given by $\mathbb{Z}_2 \times \mathbb{Z}_2 \times \mathbb{Z}_2 \times \mathbb{Z}_2$.
These four $\mathbb{Z}_2$ factors respectively correspond to the parity of the total number of electrons localized at the $1a$ Wyckoff position divided by two (i.\,e., the parity of the number of Kramers pairs), the inversion eigenvalue of the many-body wavefunction of the electrons localized at the $1a$ Wyckoff position, and the analogous two quantities evaluated for the $1b$ Wyckoff position.

\section{Chern-number constraint in the two-dimensional HK-type construction}\label{asec:chernconst}

We briefly explain the constraint used in Table~\ref{tab:inversion-eigenvalues} for the two-dimensional HK-type model. Consider a translation-invariant HK-type ground state of the form
\begin{equation}
|\Psi(L_x,L_y)\rangle
=
\bigotimes_{\bm{k}\in{\rm BZ}}
|\psi(\bm{k})\rangle ,
\end{equation}
and assume that $L_x$ and $L_y$ are both even, so that the four inversion-invariant momenta $\Gamma$, $X$, $Y$, and $M$ are present. As in the one-dimensional case, all generic momenta $\bm{k}\neq -\bm{k}$ contribute pairwise to the many-body inversion eigenvalue. Hence the many-body inversion eigenvalue is determined only by the inversion eigenvalues at the inversion-invariant momenta:
\begin{equation}
I(L_x,L_y)
=
\xi_{\Gamma}\xi_X\xi_Y\xi_M .
\end{equation}

On the other hand, for an inversion-symmetric two-dimensional wavefunction $|\psi(\bm{k})\rangle$, the product of inversion eigenvalues at the four inversion-invariant momenta gives the parity of the Chern number,
\begin{equation}
\xi_{\Gamma}\xi_X\xi_Y\xi_M
=
(-1)^{{\rm Ch}} ,
\end{equation}
as in the standard single-particle argument~\cite{Hughes-PRB-11,Fang-PRB-12}. The same argument applies here because the HK-type ground state defines a smooth momentum-space wavefunction $|\psi(\bm{k})\rangle$ as long as the gap remains open. Since time-reversal symmetry enforces ${\rm Ch}=0$, we obtain
\begin{equation}
I(L_x,L_y)=+1
\end{equation}
for even $L_x$ and $L_y$. Therefore, in contrast to the one-dimensional case, the size-independent pattern $I(L_x,L_y)=-1$ is excluded in the two-dimensional time-reversal-symmetric HK-type ground state.

For odd $L_x$ or $L_y$, not all four inversion-invariant momenta are present, and the above constraint does not apply in this form.

\bibliography{ref.bib}

\end{document}